**Hybrid Simulation of the Energy Cost of O($^1$D) and O($^3$P) Generation in a Capacitive Ar/$O_2$ Discharge Driven by Sawtooth-type Voltage Waveforms**

Jun-Xi Guo, Wan Dong*, Yuan-Hong Song*

Key Laboratory of Materials Modification by Laser, Ion and Electron Beams (Ministry of Education), School of Physics, Dalian University of Technology, Dalian 116024, People's Republic of China

*E-mail : wandong@dlut.edu.cn, songyh@dlut.edu.cn

**Abstract**

Low-pressure radio-frequency capacitively coupled plasmas (RF CCPs) operated in Ar/$O_2$ gas mixtures are widely applied in critical semiconductor manufacturing processes. O($^3$P) and O($^1$D) are key highly reactive species for oxidation or as oxygen sources for deposited thin films. Optimizing external parameters to realize efficient generation of these species under limited energy deposition is essential for improving process yield. Based on a one-dimensional (1D) fluid/electron Monte Carlo (EMC) hybrid model, this study investigates the energy cost of O($^1$D) and O($^3$P) generation driven by sawtooth up-type voltage waveforms at a fixed peak-to-peak voltage, focusing on the effects of the harmonic number ($N$) and the $O_2$ ratio. The results show that O($^3$P) generation is consistently more efficient than that of O($^1$D). The generation energy cost decreases with increasing $O_2$ ratio, yet increases as $N$ increases. The above trends are attributed to the synergistic effect of electron density, power absorption rate, and generation reaction rate coefficients. However, in the specific scenario of 10% $O_2$, an inflection point can be observed at $N$ = 2. As $N$ increases from 1 to 2, the discharge mode shifts from the DA mode to the α–DA hybrid mode, expanding the effective spatio-temporal range of the ionization rate and enhancing its peak, which increases electron density. Consequently, the generation rates are significantly enhanced, leading to a reduction in the generation energy cost. Moreover, as discussed above, monotonically increasing the harmonic number $N$ does not reduce the generation energy cost of O($^1$D) and O($^3$P) associated with medium-energy (8-20 eV) electrons. Only by selecting the appropriate $N$ to sustain the discharge in the hybrid α–DA mode, thereby increasing the electron density and promoting the generation of these species, can the generation energy cost be reduced.

## 1 Introduction

Low-pressure RF CCPs operated in $O_2$ gas are extensively employed in semiconductor manufacturing processes, such as plasma etching and thin-film deposition, as the generated atomic oxygen acts as a highly reactive oxidizing species, ion bombardment activates and modifies surfaces, or the plasma provides a continuous O source for oxidation and film growth [1-8]. However, compared with $O_2$ gas, the addition of Ar in the Ar/$O_2$ gas mixtures can be ionized to generate high-energy electrons, thus accelerating the dissociation of $O_2$ to produce more key species [9]. Therefore, it is essential to investigate the discharge characteristics of Ar/$O_2$ gas mixtures, to provide physical insight and guidance for process optimization. In thin-film deposition processes, during the formation of $SiO_2$ thin films, atomic oxygen in ground state O($^3$P), metastable atomic oxygen O($^1$D), ozone $O_3$, and metastable molecular oxygen ($O_2(a^1\Delta_g)$ 、 $O_2(b^1\Sigma_g^+)$) promote the combination of precursors with free radicals on the film surface, which serve as the main reactants in the film oxidation process [4, 10]. In particular, metastable atoms O($^1$D) and ground state atoms O($^3$P), with relatively low excitation thresholds, can lower the temperature required for thin-film deposition and play a crucial role in oxidation processes for temperature-sensitive materials. For instance, Kaspar et al [11] demonstrated that metastable atoms O($^1$D) can rapidly form a 2 nm-thick $SiO_2$ layer at a low temperature of 400ºC, and play a crucial role in the initial stage of oxidation. Moreover, Tanimura et al [12] also found that $O_2$ can reduce the impurity concentration in the film, and a higher O($^1$D) density is conducive to improving the physical and electrical properties of the film. Besides, for the formation of $B_2O_3$ thin films, Aparna et al [13] found that ground state atoms O($^3$P), due to their relatively low energy, can effectively decompose (-$OCH_3$) radicals to reduce carbon impurities in the film. For the etching process, Sankaran et al [14] demonstrated that ground state atoms O($^3$P) under an appropriate $O_2$ flow rate can effectively cleave the C-F bonds and C-H bonds in polymers, thereby removing residual polymer films after fluorocarbon plasma etching. Furthermore, Hartney et al [15] noted that ground state atoms O($^3$P) initiate chain oxidation through efficient hydrogen abstraction reactions, stripping hydrocarbon photoresists into volatile products such as CO and $CO_2$. This facilitates efficient resist stripping while maintaining the cleanliness of the etching chamber. That is, the above studies highlight the critical importance of controlling key species.

However, numerous studies on gas discharges have primarily focused on plasma characteristics, such as the transitions of discharge mode, and the control of ion energy and flux [16-49]. Based on previous studies, four discharge modes have been identified: α-mode [23, 24], γ-mode [25-28], DA-mode [29, 30], and striation (STR) mode [31]. For mixtures of electronegative and electropositive gases, particularly highly electronegative mixtures, multiple discharge modes may coexist, and transitions between them can be triggered by variations in external control parameters. For example, in capacitively coupled Ar/$O_2$ mixed gas discharges, Derzsi et al [16] demonstrated that changes in the driving frequency can induce mode transitions. Besides, in our previous studies, the effects of gas pressure and $O_2$ ratio on discharge mode transitions in single-frequency capacitively coupled Ar/$O_2$ mixed gas discharges were analyzed [32]. Furthermore, in capacitively coupled Ar/$O_2$ plasmas driven by sawtooth up-type voltage waveforms, variations in the number of harmonics, gas pressure, and $O_2$ ratio were shown to strongly affect discharge mode transitions, with the discharge evolving from a hybrid α–DA mode to a DA mode as the $O_2$ ratio increases [33]. Besides, to achieve independent control of plasma etching or deposition processes, studies on the decoupling of ion energy and ion flux are typically required. In recent years, dual-frequency (DF) driving voltage waveforms and tailored voltage waveforms (TVWs) driven by multi-frequency sources have been successively proposed in both industrial applications and fundamental research [25, 34-49]. DF driving voltage waveforms achieve this independent control by adjusting the low and high frequency voltage amplitude [34-36]. However, due to the coupling effect between the high/low frequencies, their applicability is also subject to certain limitations [37, 38]. For TVWs, there are two general types, i.e., amplitude asymmetric [39-41] and slope asymmetric voltage waveforms [42-44]. Amplitude asymmetric voltage waveforms (e.g., peak-type or valley-type voltage waveforms) can generate a self-bias voltage by adjusting the phase angle between the fundamental frequency and harmonic frequency, thus affecting the sheath properties near the electrode on both sides and ultimately realizing the modulation of the ion energy distribution [25, 45]. Meanwhile, by maintaining the driving voltage and frequency fixed, the ion flux remains essentially unchanged [25, 45, 46]. Slope asymmetric voltage waveforms (e.g., sawtooth up-type or sawtooth down-type voltage waveforms) typically regulate electron dynamic characteristics by changing the number of harmonics, thus affecting ionization/excitation reaction rates [47, 48] and ultimately realizing the modulation of ion flux, while ensuring a small variation range of self-bias voltage, thus maintaining the ion energy essentially unchanged [47-49]. Overall, tailored voltage waveforms enable effective control of ion energy and ion flux, providing greater flexibility for future nanoscale material modification.

Nevertheless, only a limited number of studies have addressed the control of neutral species flux. For example, in our previous work [32, 33], a detailed analysis of the evolution of charged and neutral species with changes in discharge parameters under both single-frequency and sawtooth up-type voltage waveforms is conducted. Besides, to quantitatively analyze how to regulate external parameters for the efficient generation of target neutral species, Wang et al [50] defined the energy cost of the generation of F atoms (denoted as $\eta$) in low pressure (4 Pa) $CF_4$ CCPs, which is the ratio of the total power absorption rate of charged particles to the total F atom generation rate, representing the energy required to generate a single F atom. By using a PIC/MCC model with a stationary neutral-species diffusion model, they further investigated the influence of different driving voltage waveforms on the energy cost of F atoms generation and demonstrated independent control of ion energy and radical density. But, systematic investigations into the energy cost of key neutral-species generation remain limited in $Ar/O_2$ mixed gas discharges.

Therefore, this study will conduct a detailed investigation of the generation energy cost of key neutral species, and generation/loss mechanism in capacitively coupled $Ar/O_2$ mixed gas discharges. Based on the one-dimensional (1D) fluid/electron Monte Carlo hybrid simulations, and adopting the definition of neutral-species generation energy cost proposed by Wang et al. [50], the regulation mechanisms of the number of harmonics and $O_2$ ratio on electron dynamics are clarified, together with their impacts on the power deposited into the plasma and the generation rates of $O(^1D)$ and $O(^3P)$ atoms. On this basis, the correlation with the generation energy cost of the target neutral species, $O(^3P)$, $O(^1D)$ is established, providing a theoretical reference for practical plasma processes while reducing their generation energy cost. The structure of the article is as follows: Section 2 briefly introduces the simulation model, Section 3 delves into the results' analysis, and the final section presents the conclusions.

## 2 Model description

In this study, a one-dimensional (1D) fluid coupled with electron Monte Carlo (1D/3V) hybrid model is used to investigate the discharge characteristics of capacitively coupled $Ar/O_2$ mixed gas discharges driven by sawtooth up-type voltage waveforms with a fundamental frequency of 13.56 MHz, and analyzes the generation energy cost of $O(^3P)$ and $O(^1D)$ atoms as a function of the number of harmonics and $O_2$ ratio.

In the fluid model, the main equations include the continuity equations for electrons and ions,

the electron drift-diffusion equation, the electron energy balance equation, the ion momentum balance equation, the neutrals continuity equations coupled with diffusion, and Poisson's equation. The temperatures of ions and neutral species are assumed to be equal to that of the background gas; therefore, their energy balance equations are not solved. Detailed descriptions of these equations can be found in our previous studies [32, 33, 51-53]. The fluid model provides the temporal and spatial distributions of electron, ion, and neutral densities, together with the corresponding electric field characteristics. Regarding the boundary conditions at the electrodes, the electron density ($n_e$) is assumed to follow the Boltzmann distribution. The electron flux is expressed as $\Gamma_{e,x|x=0,d} = \mp \frac{1}{4} n_e u_{th,e}(1\text{-}\Theta) - \gamma \Gamma_{i,x|x=0,d}$, where $u_{th,e} = \sqrt{\frac{8kT_e}{\pi m_e}}$ is the electron thermal velocity, and the electron reflection probability ($\Theta$) is fixed at 0.25 [54-56]. The secondary electron emission coefficient ($\gamma$) is set as zero, since it is known not to play an important role under the voltage and pressure conditions used in this work [57, 58]. The electron energy flux is given by $\Gamma_{w,x|x=0,d} = \frac{5}{2} T_e \Gamma_e$. For positive ions, the densities and fluxes are considered to be continuous, and for negative ions, the densities and fluxes are set to zero. The neutral density ($n_n$) is considered to be continuous, and the neutral flux is set as $\Gamma_{n,x|x=0,d} = \frac{s_i}{2(2\text{-}s_i)} n_n u_{th,n}$, where $u_{th,n}$ and $s_i$ represent the neutral thermal velocity and viscosity coefficient. The diffusion coefficient of neutrals ($D_n$) is considered based on the Blanc law [59], which is determined by the corresponding coefficients in the background gas Ar and $O_2$. Detailed expressions for $D_n$, along with the binary collision diameters ($\sigma_i$), potential energies ($\varepsilon_i/k_b$), and viscosity coefficients ($\eta_i$) of neutrals, as well as other relevant coefficients, can be found in our previous work [32, 33, 60].

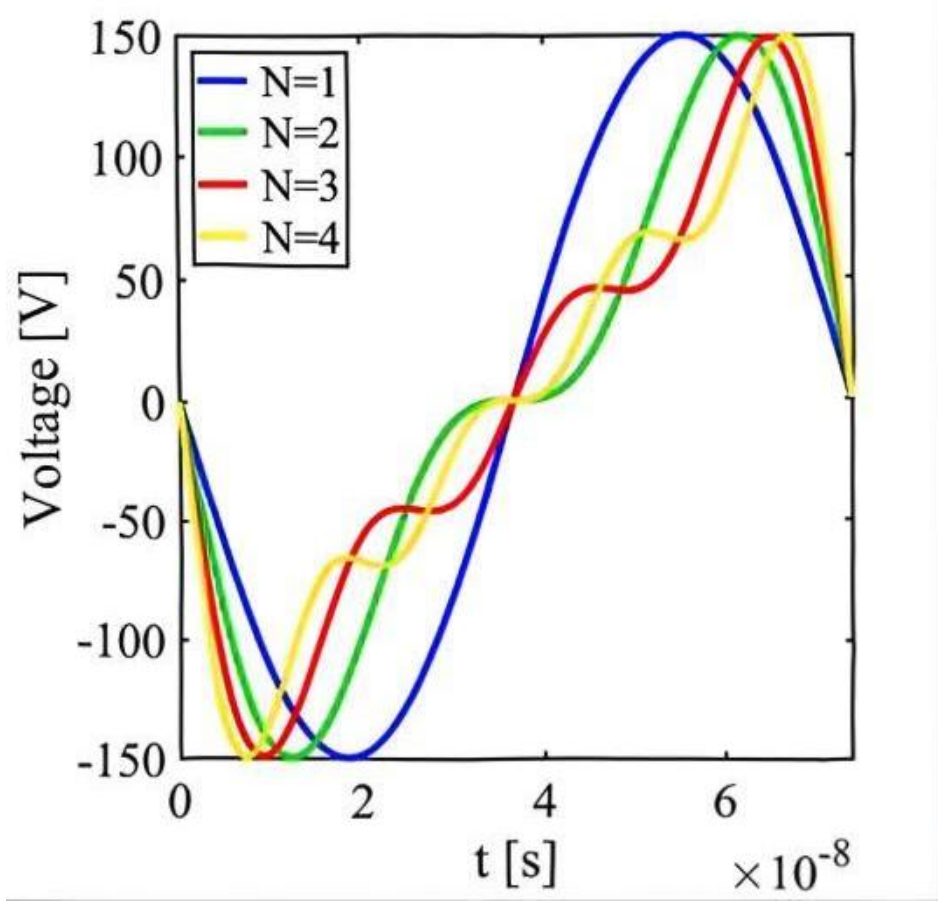

**Figure 1** The sawtooth up-type voltage waveforms: $N = 1\text{-}4$, fundamental frequency $f$ of 13.56 MHz, and peak-to-peak voltage $V_{pp}$ of 300 V.

The electric potentials at the powered electrode ($x = 0$) and the grounded electrode ($x = d$) satisfy $V|_{x=0} = V(t)+V_{DC}$ and $V|_{x=d} = 0$ respectively. The sawtooth up-type voltage waveform is given by $V(t) = V_a \sum_{k=1}^{N} \frac{1}{k} sin(2\pi kft)$, where the fundamental frequency $f$ is 13.56MHz, and the voltage amplitude $V_a$ is set to ensure a peak-to-peak value of 300 V. As shown in figure 1, the voltage waveform maintains a constant amplitude while exhibiting slope asymmetry—with the waveform fall time accelerating and rise time decelerating as $N$ increases. This induces

non-sinusoidal motion of the sheaths on both sides, thereby leading to the electric asymmetry effect (EAE). For simplicity, only the cases for $N$ = 1-4 are presented. The DC self-bias ($V_{DC}$) induced by EAE is calculated by the time-averaged flux balance of positive ($Q_p$) and negative ($Q_n$) particles on the powered electrode, with the iterative update rule defined as follows: $Q_p > Q_n$, $V_{DC}^{t+1} = V_{DC}^{t} + \Delta V$; $Q_p < Q_n$, $V_{DC}^{t+1} = V_{DC}^{t} - \Delta V$, where $\Delta V$ is a small increment, typically set to 0.01 V. The iterative process is continued until current balance is achieved.

In the electron Monte Carlo model, 30,000 pseudoparticles are employed to represent electrons in the computational domain in order to obtain statistically well-resolved electron energy distributions and related electron properties. According to Newton's second law, electrons move under the influence of the electric field while electron-neutral collisions are taken into account. Electron dynamic variables are recorded at each time step, and the electron energy distribution function (EEDF) is obtained statistically after five cycles of the fundamental frequency. Further details of the model can be found in our previous studies [32, 33, 51-53, 61, 62].

For coupling the two models, the fluid model is first operated for only the initial five fundamental-frequency cycles to generate a preliminary spatio-temporally evolving electric field. This field data are stored and transferred to the electron Monte Carlo model, where the spatio and temporal evolution of EEDF is obtained statistically. Based on the EEDF, together with auxiliary information such as collision reaction cross sections, key parameters—including the electron temperature, reaction rate coefficients, and electron transport coefficients—are calculated; detailed formulations can be found in references [53, 61, 62]. These parameters are then fed back into the fluid model. In the subsequent stage, the electron energy balance equation in the fluid model is no longer solved; instead, the electron temperature evaluated from the electron Monte Carlo model is employed. After each execution of the fluid model, the updated electric field is again transferred to the electron Monte Carlo model, forming an iterative loop. This loop is continued for approximately 60000 fundamental-frequency cycles until a steady state is reached. The steady state is defined as a condition in which all calculated variables vary only within a single fundamental-frequency cycle and differ from those in the preceding cycle by less than 0.1%. The resulting data are then used for further analysis. In addition, owing to the distinct time scales of neutral species compared with electrons and ions, a larger time step is adopted for neutrals to reduce the overall computational cost.

In the $Ar/O_2$ reaction set, five charged species ($e^-$, $Ar^+$, $O_2^+$, $O^+$, and $O^-$) and four neutral species ($O(^3P)$, $O(^1D)$, $O_2(a^1\Delta_g)$, and $Ar^*$) are included. The reaction set adopted in this work is based on our previous studies of $Ar/O_2$ gas mixtures [32, 33] and comprises a total of 56 collision processes, including elastic scattering, ionization, excitation, and dissociative attachment reactions. In addition to electron-neutral interactions, ion-ion, ion-neutral, and neutral-neutral collisions are also taken into account. Moreover, the hybrid model employed in this study has been experimentally validated, as reported in reference [33].

## 3 Results and discussions

In this section, a comprehensive analysis is conducted on the plasma power absorption mechanisms and the generation mechanisms of the key neutral species, $O(^1D)$ and $O(^3P)$ atoms. Furthermore, the generation energy cost of these key neutral species, $O(^1D)$ and $O(^3P)$, is evaluated under various external operating parameters. The discharge conditions are specified as follows: a fundamental frequency of 13.56 MHz, a peak-to-peak voltage of 300 V, a gas pressure of 200 mTorr, and a background gas temperature of 400 K. The electrodes are arranged in a parallel configuration with a gap of 3 cm. The effects of the number of harmonics ($N$ = 1-6) and the $O_2$ ratio ($r$ = 10%, 30%, and 50%) are investigated separately. In this paper, the variable $r$ is defined as the percentage of background gas $O_2$ in the Ar/$O_2$ gas mixtures, i.e., $r = \frac{[O_2]}{([Ar]+[O_2])} \times 100\%$.

Figure 2 shows the spatially and temporally averaged generation energy cost of O($^1$D) and O($^3$P) atoms under different number of harmonics $N$ and $O_2$ ratio $r$. Following the definition method proposed by Wang et al. [50], the energy cost of the generation of neutral species is defined as the ratio of the total power absorption rate of charged particles to the total generation rate of the neutral species, i.e., the energy required to generate one neutral particle. In contrast, the energy efficiency is inversely proportional to the energy cost, referring to the number of neutral species generated per unit energy. The relevant expression of the generation energy cost is given as follows: $\eta = \frac{P_{total}}{S_{gtotal}}$, where $\eta$ means the energy cost of the generation of neutral species with the unit of eV per atom, $S_{gtotal}$ is the sum of the spatially and temporally averaged rates of all generation reactions for corresponding neutral species, $P_{total} = \sum_j P_j = \sum_j J_j \cdot E$ is the spatially and temporally averaged total power absorption rate of electrons and positive ions. For power absorption rate ($P_j = J_j \cdot E$), it is obtained based on current density ($J_j$) and the electric field ($E$), where $j$ denotes the charged particle species.

For metastable atoms O($^1$D), according to the reaction set described in the model and previous studies [32, 33], their generation primarily depends on electron-induced dissociative collisions. The dominant reaction pathway is e + $O_2$ → O($^3$P) + O($^1$D) + e (R1), with a dissociation threshold energy of 8.4 eV. Besides, as illustrated in figure 2 (a), as $r$ increases, the energy cost for generating metastable atom O($^1$D) decreases continuously. In other words, the energy required to generate a single metastable O($^1$D) atom is progressively reduced, corresponding to an enhancement in generation energy efficiency. This behavior is primarily attributed to the significant increase in the background $O_2$ density with increasing $r$, which enhances the rate of reaction R1 and thereby promotes the generation of O($^1$D). Consequently, the total generation rate of O($^1$D) increases faster than the total power absorption rate of charged particles, leading to a continuous improvement in generation energy efficiency. At relatively higher $O_2$ ratio ($r$ = 30%, 50%), the energy cost of generating metastable atom O($^1$D) increases monotonically with increasing $N$, while the energy efficiency decreases continuously. In contrast, at lower $O_2$ ratio ($r$ = 10%), the generation energy cost does not exhibit a monotonic trend but instead shows an overall increasing tendency with an inflection point at $N$ = 2, corresponding to an initial rise followed by a decline in energy efficiency.

For ground state atoms O($^3$P), as illustrated in figure 2 (b), the variation trend of the generation energy cost with respect to $r$ and $N$ is nearly consistent with that of metastable O($^1$D) atoms, although a quantitative difference is observed. For example, under identical discharge conditions, the generation energy cost of O($^3$P) atoms is consistently lower than that of O($^1$D) atoms by approximately 10–15 eV per atom. This is because the generation pathways of O($^3$P) atoms involve not only electron-induced dissociative collisions but also neutral-neutral collisions. At low $O_2$ ratios ($r$ = 10%), the dominant reactions include e + $O_2$ → O($^3$P) + O($^1$D) + e (R1), e + $O_2$ → O($^3$P) + O($^3$P) + e (R2) and $O_2$ + $Ar^*$ → O($^3$P) + O($^3$P) + Ar (R3), where the dissociation threshold energy of reaction R2 is 6.0 eV. However, as the $O_2$ ratio increases ($r$ > 10%), the primary neutral-neutral collision shifts from reaction R3 to $O(^1D) + O_2 \rightarrow O(^3P) + O_2$ (R4), making reactions R1, R2, and R4 the dominant reaction pathways. It can be clearly observed that, owing to the multiple generation pathways contributing to $O(^3P)$, its total generation rate surpasses that of $O(^1D)$, and further leading to a lower generation energy cost of $O(^3P)$. In addition, under the same $O_2$ ratio and different number of harmonics $N$, the variation amplitude of the $O(^3P)$ generation energy cost is weaker than that of $O(^1D)$. For example, as the number of harmonics increases from $N$ = 1 to 6 at an $O_2$ ratio of 50%, the difference between the maximum and minimum generation energy cost is approximately 8 eV for $O(^1D)$ and about 2.5 eV for $O(^3P)$. This can also be attributed, to some extent, to the contribution of neutral-neutral collision reactions (R3 and R4), which reduces the sensitivity of the generation process to electron dynamics.

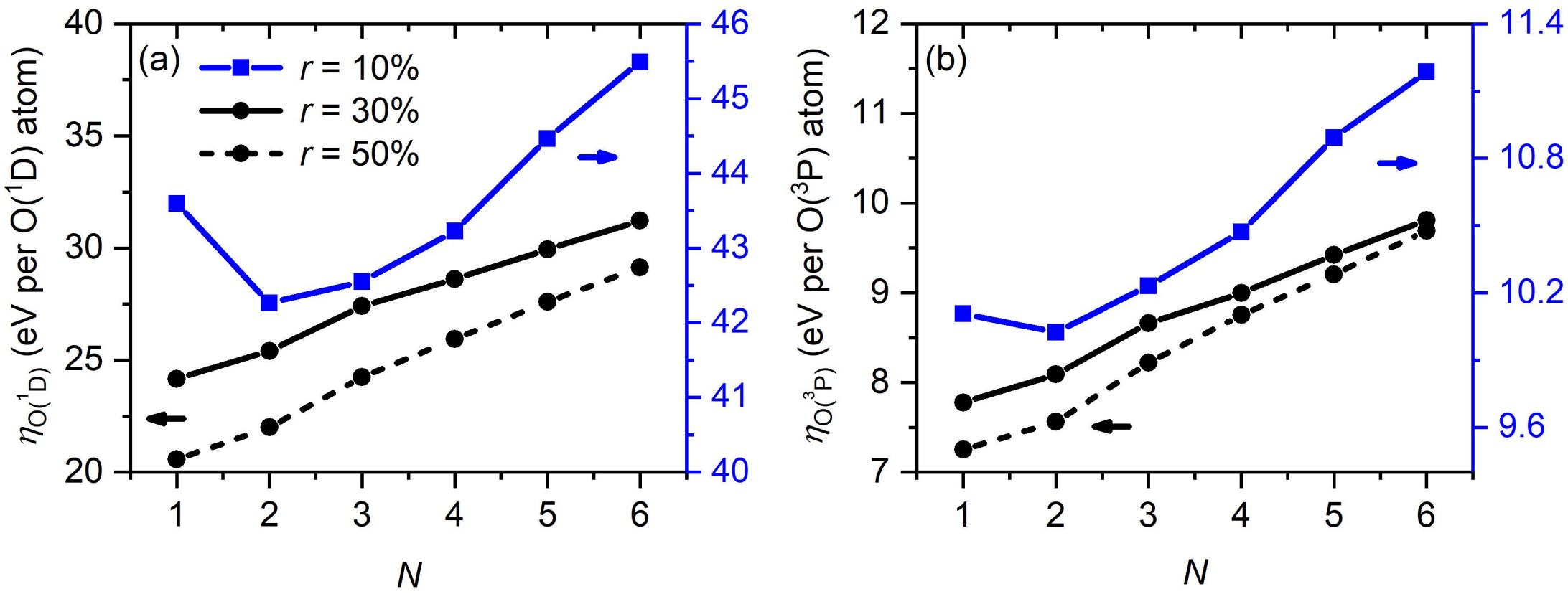

**Figure 2** Spatially and temporally averaged generation energy cost of metastable atom $O(^1D)$ (a), ground state atom $O(^3P)$ (b) as a function of $O_2$ ratio ($r$ = 10%, 30%, 50%) and number of harmonics ($N$ = 1-6). Discharge conditions: The discharges is operated at a fundamental frequency of 13.56 MHz, a peak-to-peak voltage of 300 V, a gas pressure of 200 mTorr, and a background gas temperature of 400 K. The electrodes are arranged in a parallel configuration with a gap of 3 cm.

To better understand the variation in the generation energy cost of these neutral species, particularly the inflection point observed at an $O_2$ ratio of 10%, the total power absorption rate of charged particles and the total generation rates of these neutral species will be systematically analyzed. In addition, since the generation energy cost of $O(^1D)$ and $O(^3P)$ atoms exhibits similar variation trends at relatively high $r$ values ($r$ = 30% and 50%), the results at $r$ = 50% are selected as representative of highly electronegative conditions. These are then compared with the results at $r$ = 10%, which represent weakly electronegative conditions.

Figure 3 shows the spatially and temporally averaged power absorption rate of charged particles and the total generation rates of $O(^1D)$ and $O(^3P)$ atoms at different number of harmonics $N$ and $O_2$ ratio $r$. Figures 4 and 5 present the spatio-temporal evolutions of the electron density, electron power absorption rate, total ionization rate, main generation reaction rate of $O(^1D)$, the sum of the main generation reaction rates for $O(^3P)$, and the $O^-$ density at $O_2$ ratios $r$ of 10% and 50%, with the number of harmonics $N$ varying from 1 to 4. Figure 3 (a) shows that, under the pressure condition employed in this study (200 mTorr), the ion power absorption rate is consistently lower than that of electrons and remains nearly independent of both the number of harmonics $N$ and the $O_2$ ratio $r$. This can be attributed primarily to the fact that the gas pressure, which plays a crucial role in ion heating, is maintained constant. Consequently, the ion energy loss due to ion–neutral charge exchange collisions in the sheath remains nearly unchanged under different discharge conditions. Furthermore, the sawtooth up-type voltage waveform mainly modifies the sheath dynamics (and thus the electron dynamics), while having only a limited effect on the DC self-bias voltage within the sheath. Specifically, the self-bias voltage is at a low level, and its amplitude decreases slightly with increasing $N$ (from -10.54 V to -5.6 V at $r$ = 10%; from -29.39 V to -18.84 V at $r$ = 50%). As a result, ion heating remains weak, showing only minor variations across different discharge parameters.

In contrast, the electron power absorption rate exhibits a pronounced increase with increasing harmonic number $N$ and $O_2$ ratio $r$, except for a minimum observed at $N$ = 2 with an $O_2$ ratio of 50%. The total power absorption rate follows a similar dependence to that of the electron power absorption rate, differing only in magnitude. Besides, the electron power absorption rate increases continuously with increasing $r$. This behavior can be primarily attributed to the enhancement of electron heating in the bulk region. As $O_2$ ratio $r$ increases, the spatially and temporally averaged electronegativity is enhanced (e.g., from 5.73 to 12.50 at $N$ = 1), while the electron density in the bulk region decreases (see figures 4 (a1–a4) and 5 (a1–a4)), leading to a reduction in electrical conductivity. Consequently, both the drift electric field in the bulk and the ambipolar electric field near the sheath edge are significantly strengthened. For example, at $N$ = 1 and $t = 0.1T_{RF}$, the spatially averaged electric field near the sheath region increases from −7.02 V/cm to −11.07 V/cm, thereby enhancing electron heating.

To quantitatively characterize the relative variation of physical quantities between adjacent number of harmonics $N$, the consecutive relative growth rate is defined as $s_{i+1} = \frac{X_{i+1}-X_i}{X_i} \times 100\%$, where $s_{i+1}$ denotes the consecutive relative growth rate of the physical quantity from $N = i$ to $N = i+1$, $X_i$ and $X_{i+1}$ are the magnitudes of the target physical quantity at $N = i$ and $N = i+1$. In this work, this definition will be applied to the power absorption rate, generation energy cost, and the generation rate of neutral species. When the $O_2$ ratio is fixed at 10%, the electron power absorption rate increases as $N$ increases. Specifically, with increasing $N$, the power absorption rate increases with an average consecutive relative growth rate of 9.38%. However, at $N$ = 2, the consecutive relative growth rate is relatively small, only 5.28%. As the number of harmonics $N$ increases from 1 to 6, the slope asymmetry effect of the sawtooth up-type voltage waveform causes the sheath near the powered electrode to exhibit a motion characteristic of rapid expansion and slow collapse. Besides, the increase in $N$ continuously accelerates the sheath expansion rate, leading to a significant enhancement in the electric field amplitude at the sheath expansion edge (from -4.54 V/cm to -13.27 V/cm at $t = 0.1T_{RF}$), ultimately leading to enhanced electron heating, as shown in figures 4 (b1-b4). Nevertheless, at $N$ = 2 (see figure 4 (b2)), in the bulk region, although the peak of power absorption rate increases slightly compared with $N$ = 1, the pronounced asymmetry of the sheath motion significantly shortens its effective action time. Moreover, at the sheath edge near the powered electrode, the sheath expansion rate at $N$ = 2 remains relatively low, leading to only a small difference between the positive and negative power absorption rates. After temporal averaging, the enhancement in power absorption rate is therefore mitigated, resulting in a relatively modest increase at $N = 2$.

When $r = 50\%$, with increasing $N$, the electron power absorption rate decreases first and then increases, reaching the minimum at $N = 2$ (its consecutive relative growth rate is -6.73%). The average consecutive relative growth rate is 9.42% for $N$ = 3-6. It can be observed that the phenomenon of a reduced or even negative consecutive relative growth rate becomes more pronounced at $N = 2$ with higher $O_2$ ratio. At $N = 2$, compared with the case of $N > 2$, the sheath expansion rate is still relatively slow, and, the electron density at the sheath edge on the powered electrode side decreases (see figure 5 (a2)), leading to a slight decrease in power absorption rate at $N = 2$ (see figure 5 (b2)). As $N$ continues to increase, the sheath expansion rate rises significantly, and the electron density increases, causing the power absorption rate gradually increases continuously (see figures 5 (b3–b4)).

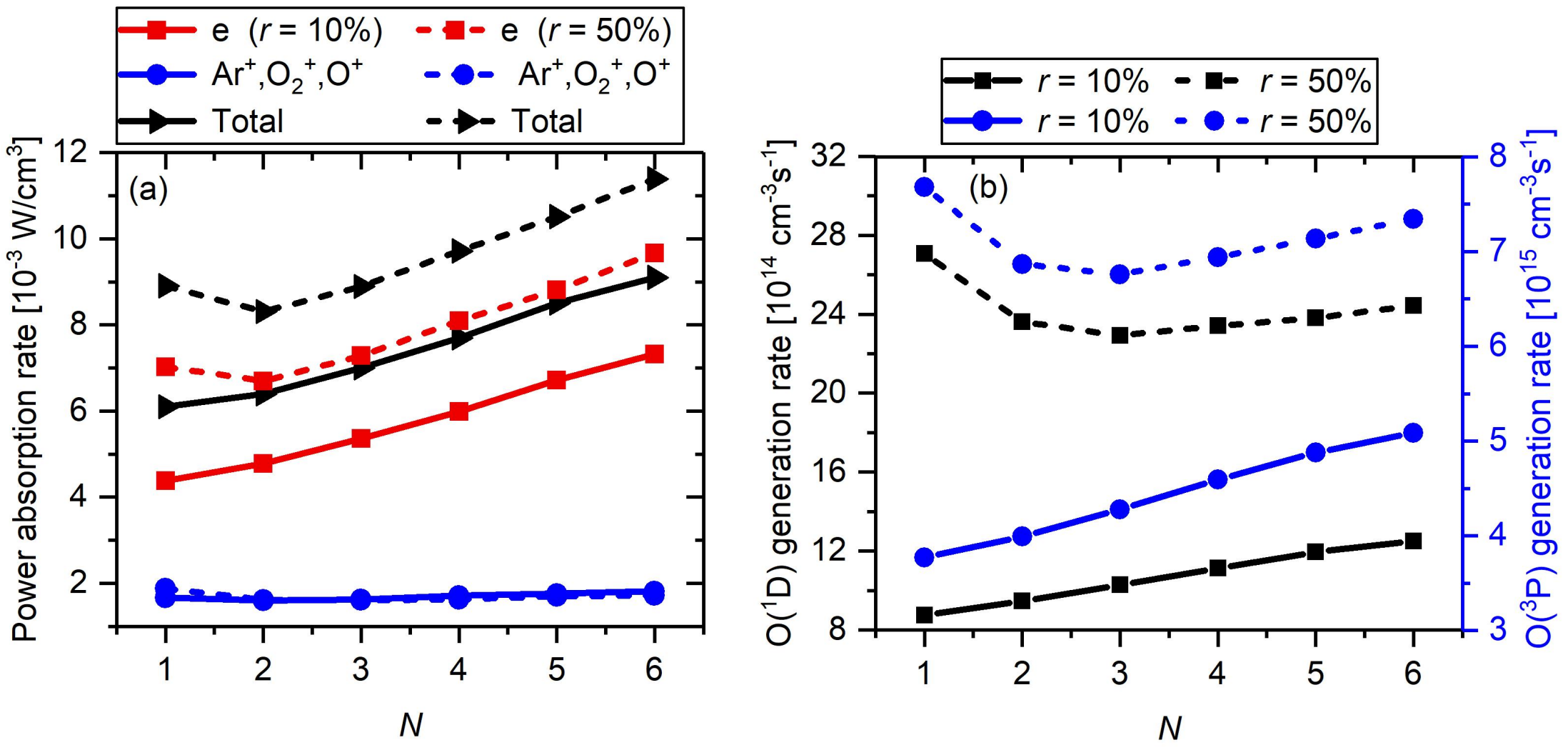

**Figure 3** Spatially and temporally averaged electron, positive ion, and the total power absorption rate (a), and total generation rates of metastable atom $O(^1D)$ and ground state atom $O(^3P)$ (b) at different $O_2$ ratio $r = 10\%$ (solid line), 50% (dotted line) and number of harmonics ($N = 1$-6). The discharge conditions are consistent with those shown in figure 2.

As shown in figure 3 (b), with increasing $r$, the total generation rate of $O(^1D)$, Gen_O($^1$D), increases, which is directly related to the increase in the background $O_2$ density. Moreover, for $r = 10\%$, as $N$ increases, Gen_O($^1$D) increases monotonically. As shown in figures 4 (d1–d4), as $N$ increases, the peak of the generation rate increases significantly (from $1.49\times10^{15}$ cm$^{-3}$s$^{-1}$ to $2.62\times10^{15}$ cm$^{-3}$s$^{-1}$), while the region of effective action time and space remains nearly unchanged. Consequently, the spatially and temporally averaged total generation rate, Gen_O($^1$D), increases as a function of $N$.

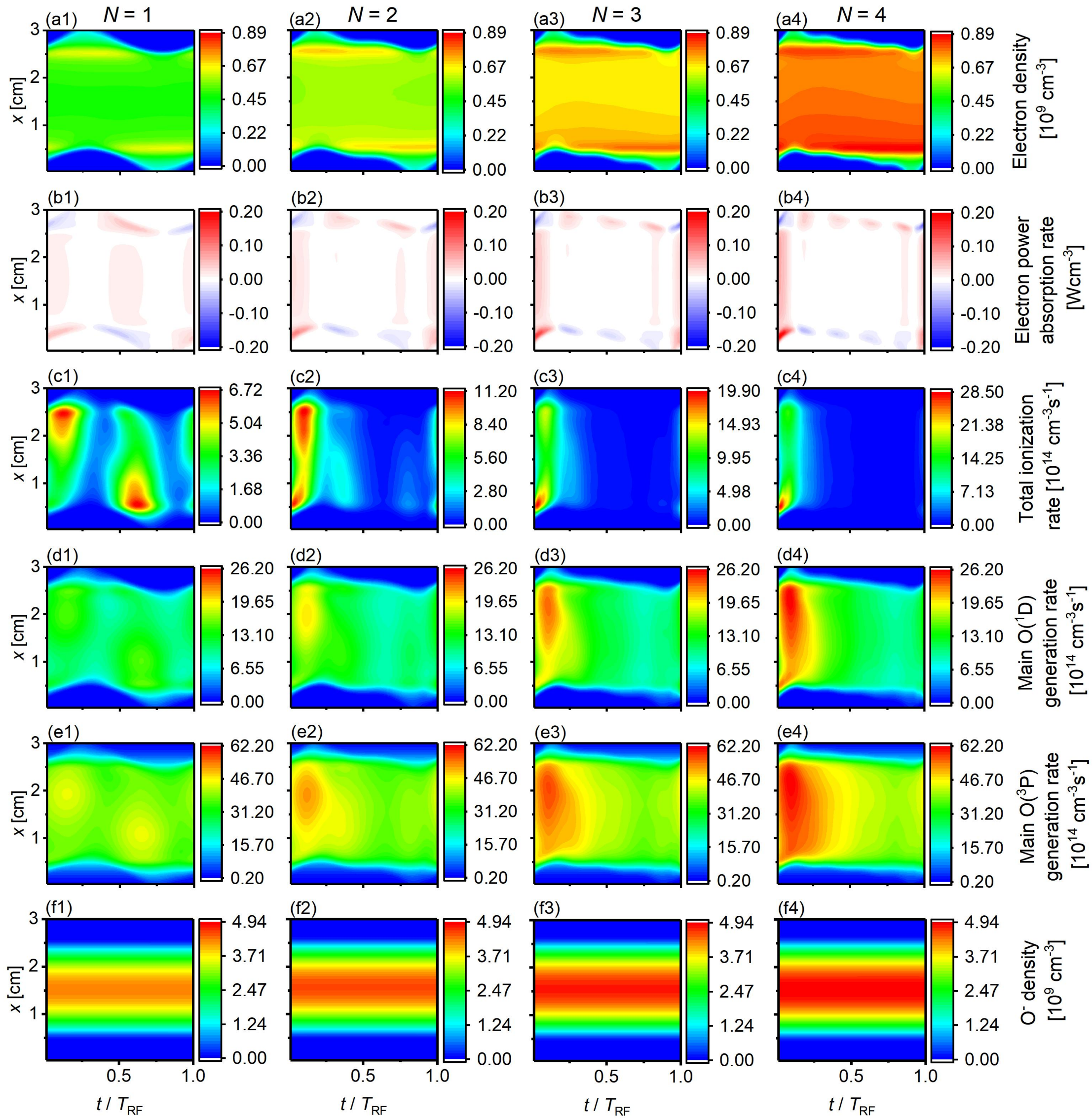

**Figure 4** Spatio-temporal evolutions of electron density (a1)-(a4), electron power absorption rate (b1)-(b4), total ionization rate (c1)-(c4), main generation reaction (R1) rate of metastable atom $O(^1D)$ (d1)-(d4), the sum of main generation reaction (R1, R2, R3) rates for ground state atom $O(^3P)$ (e1)-(e4), and $O^-$ density (f1)-(f4) at $O_2$ ratio ($r$ = 10%) and number of harmonics ($N$ = 1-4). The discharge conditions are consistent with those shown in figure 2.

For $r = 50\%$, Gen_O($^1$D) initially decreases and then increases slightly with increasing $N$, reaching its minimum at $N$ = 3. As illustrated in figures 5 (d1-d4), the generation rate is mainly localized near the sheath collapse phase near the grounded electrode in the case of $r$ = 50%. Besides, the peak of the generation rate increases continuously with increasing $N$ (from $6.50\times10^{15}$ cm$^{-3}$s$^{-1}$ to $12.75\times10^{15}$ cm$^{-3}$s$^{-1}$), while the region of effective action time and space decreases. The combined effect of the above two factors determines the variation trend of Gen_O($^1$D). When $N \leq 3$, the decrease in the region of effective action time and space of the generation rate is the dominant factor, resulting in a reduction of Gen_O($^1$D). However, as $N$ further increases ($N > 3$), the contraction of the region of effective action time and space becomes less pronounced, and the growth of its peak value consequently leads to an increase in Gen_O($^1$D). Besides, Gen_O($^1$D) reaches a minimum at $N = 3$.

From figure 3 (b), the overall variation trend of ground state atoms $O(^3P)$ is consistent with that of metastable atoms $O(^1D)$, with their corresponding values being numerically higher. This is because, as shown in figures 4 (e1–e4) and figures 5 (e1–e4), their spatio-temporal evolutions are similar to those of metastable atoms $O(^1D)$ but feature a longer effective action time. Furthermore, as mentioned above, this is attributed to the fact that the primary generation pathways of $O(^3P)$ involve not only R1 but also R2, R3, and R4.

Although the variation trends of the generation rates of $O(^1D)$ and $O(^3P)$ have been shown above, the underlying factors governing these trends still require further investigation. Since the total generation rates (Gen_$O(^3P)$, Gen_$O(^1D)$) are related to the reaction rate coefficient $k$, background gas density and electron density $n_e$. Since the background gas density is kept constant, the following discussion focuses on the effects of variations in the $k$ and $n_e$ on the total generation rates. The electron-induced collision reaction rate coefficient $k$ is mainly determined by the collision cross section of the relevant reactions and the EEDF. Therefore, figures 6 and 7 show collision reaction cross sections between electrons and $O_2$ molecules in the main generation reactions of metastable atom $O(^1D)$ and ground state atom $O(^3P)$, and spatially and temporally averaged EEDF at different number of harmonics $N$. And, tables 1 and 2 show the consecutive relative growth rates of spatially and temporally averaged electron density $n_e$, rate coefficient of reaction R1, $k$(R1), and rate coefficient of reaction R2, $k$(R2), as a function of $N$ with the $O_2$ ratios of 10% and 50%, respectively.

For metastable atoms $O(^1D)$, the collision cross sections shown in figure 6 indicate that the primary generation reaction, R1, depends on electrons with energies in the range of 10-20 eV as well as those above 20 eV. However, as illustrated in figure 7, electrons with energies exceeding 20 eV are confined to the high-energy tail of the EEDF and account for only a small fraction of the total electron population. Consequently, the generation of metastable $O(^1D)$ atoms is particularly sensitive to electrons in the 10-20 eV energy range.

From table 1, for $r = 10\%$, the consecutive relative growth rate of $n_e$ decreases from 18.60% to 8.54%, and the consecutive relative growth rate of $k$(R1) increases from -8.74% to -2.56% with increasing $N$ (2-6). Regarding the increase in electron density, it is because the rapid expansion of sheath near the powered electrode causes a continuous rise in the peak of the total ionization rate (see figures 4 (c1-c4)), and further causing the increasing of electron density $n_e$. Especially at $N = 2$, the rapid expansion of the sheath near the powered electrode also enhances the α-mode discharge, with the discharge intensities of the DA-mode and α-mode being comparable. This is manifested by ionization rate peaks distributed at the sheath edges of both electrodes as well as in the bulk plasma region (see figure 4 (c2)), showing a pronounced increase in the effective spatio-temporal region. This leads to the consecutive relative growth rate of $n_e$ reaching a maximum value of 18.60% at $N = 2$. When $N > 2$, as $N$ increases further, the sheath expansion velocity near the powered electrode continues to accelerate. This causes the discharge progressively transitions to being dominated by the α-mode. Although the ionization rate peaks continue to intensify, its spatial distribution becomes more localized, mainly concentrated during the sheath expansion phase near the powered electrode (see figures 4 (c3–c4)). Meanwhile, the dissociative electron attachment reaction is strengthened, leading to a gradual increase in $O^-$ ion density (see figures 4 (f3–f4)), which in turn causes a relatively slow consecutive relative growth rate of electron density $n_e$ (see table 1).

Furthermore, an enhancement of the high-energy tail of the EEDF (see figure 7) can be observed. Meanwhile, the enhancement of the high-energy tail is accompanied by a corresponding reduction in the proportion of electrons within the 10-20 eV energy range (see figure 7), leading to a continuous decrease in the *k*(R1). From the above analysis, it can be found that although *k*(R1) exhibits a negative consecutive relative growth rate, the sustained and relatively large positive increase in electron density $n_e$ results in a monotonic increase in Gen_O($^1$D), with an average consecutive relative growth rate of 7.11%, reaching a maximum value of 9.20% at $N = 2$.

Furthermore, as shown in figure 3, at $r = 10\%$, although both the total power absorption rate and the total O($^1$D) generation rate increase, the consecutive relative growth rate of the total power absorption rate reaches a minimum at $N = 2$ (5.28%). In contrast, the maximum consecutive relative growth rate of the total O($^1$D) generation rate (9.20%), primarily driven by a significant rise in electron density $n_e$, plays a dominant role on generation energy cost of O($^1$D) at $N = 2$. Consequently, the energy cost of O($^1$D) reaches a minimum at $N = 2$ (see figure 2 (a)). However, as $N$ increases from 2 to 6, the average consecutive relative growth rate of the total power absorption rate (9.38%) begins to outpace that of the total O($^1$D) generation rate (7.11%), thus the energy cost of O($^1$D) exhibits an increasing trend, indicating a decline in energy efficiency (see figure 2(a)).

This behavior contrasts with the findings of Wang et al. [50], who reported that the energy cost of F atom generation in $CF_4$ plasmas decreases monotonically with increasing $N$. This discrepancy mainly arises from the different electron energy ranges required for neutral species generation: F atom generation predominantly depends on electrons in the high-energy tail of the EEDF. Consequently, the distinct evolution of specific electron energy groups leads to opposite trends in generation energy cost as $N$ increases.

Nevertheless, based on this work, it can be found that by adjusting the number of harmonics, the discharge can operate in a hybrid α–DA mode. Meanwhile, the spatio-temporal effective region of the ionization rate is significantly expanded, accompanied by enhanced peaks, which, to some extent, increases the electron density and further plays an important role in reducing the energy cost of neutral species associated with medium-energy (10-20 eV) electrons.

For $r = 50\%$, table 2 shows that the consecutive relative growth rate of $n_e$ first increases from 9.25% to 11.99%, and then decreases to 11.22%, 8.97%, and 7.43% as $N$ increases from 2 to 6. Meanwhile, the consecutive relative growth rate of *k*(R1) increases from –21.41% to –5.57% with the increase in $N$ (2-6). Besides, from figures 5 (c1)-(c4), the discharge is dominated by DA-mode due to higher electronegativity. The peak of ionization rate is enhanced, but consistently localized during the sheath collapse phase near the grounded electrode. Furthermore, the spatio-temporal effective region of the ionization rate is reduced compared with that at $N = 1$. Meanwhile, the consumption caused by the dissociative electron attachment reaction is not pronounced. As shown in figures 5 (f2)-(f4), the $O^-$ density peak is increased from $5.93\times10^9$ cm$^{-3}$ to only $6.38\times10^9$ cm$^{-3}$. The above factors lead to a slight increase in $n_e$, but its consecutive relative growth rate remains at around 9%, with no particularly significant changes. In contrast, due to the enhanced sheath expansion field near the sheath edge, the electrons can be sufficiently accelerated. Thus, the high energy tail of the EEDF is significantly enhanced, resulting in a substantial decrease in the proportion of electrons in the 10-20 eV energy range (see figure 7 (b)). Consequently, the consecutive relative growth rate of *k*(R1) decreases markedly (particularly at $N = 2$, where it reaches −21.41%), which dominates the decline in the consecutive relative growth rate of Gen_O($^1$D). Therefore, the consecutive relative growth rate attains its minimum value of −12.92% at $N = 2$, while over the range $N = 3$–6, the average consecutive relative growth rate is 0.86%.

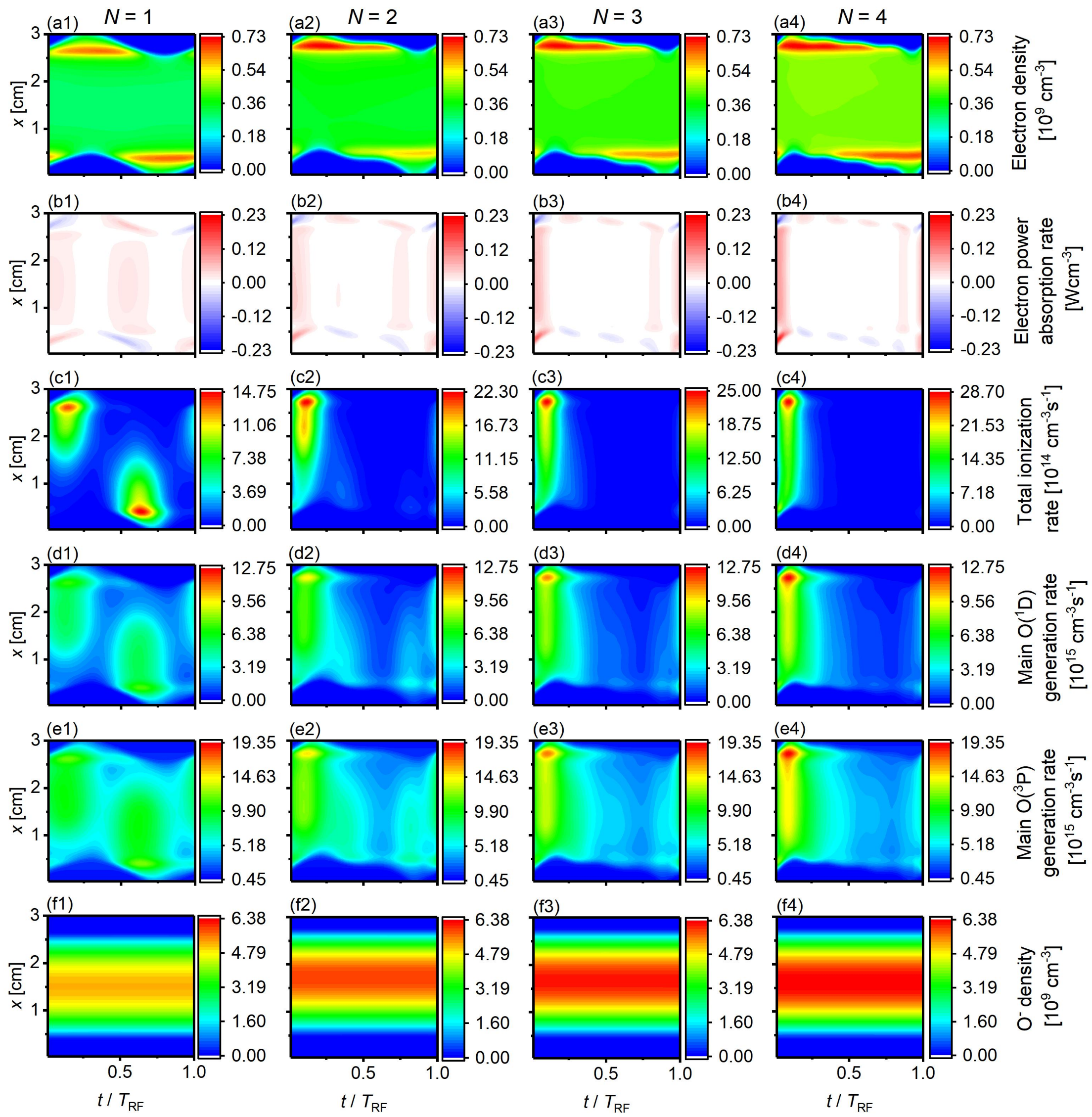

**Figure 5** Spatio-temporal evolutions of electron density (a1)-(a4), electron power absorption rate (b1)-(b4), total ionization rate (c1)-(c4), main generation reaction (R1) rate of metastable atom O($^1$D) (d1)-(d4), the sum of main generation reaction (R1, R2, R4) rates for ground state atom O($^3$P) (e1)-(e4), and $O^-$ density (f1)-(f4) at $O_2$ ratio ($r = 50\%$) and number of harmonics ($N = 1$-4). The discharge conditions are consistent with those shown in figure 2.

With increasing $N$, the total power absorption rate decreases first and then increases, reaching the minimum at $N = 2$, where the consecutive relative growth rate of power absorption rate is -6.73%. However, because the consecutive relative growth rate of the total O($^1$D) generation rate (-12.92% primarily driven by a significant decrease in $k$(R1)), is reduced significantly at $N = 2$, the generation energy cost of O($^1$D) exhibits an increasing trend compared to $N = 1$. Furthermore, the average consecutive relative growth rate of the power absorption rate (9.42%) consistently outpaces that of the generation rate (0.86%). Ultimately, the generation energy cost increases monotonically and the energy efficiency decreases continuously, with no inflection point observed.

It can be seen that at a higher $O_2$ ratio with relatively strong electronegativity and the discharge remains dominated by the DA mode. And as the number of harmonics $N$ increases, and the occurrence of a discharge governed by an α–DA hybrid mode, with an increase in the ionization peak, is relatively difficult. Therefore, the electron density $n_e$ cannot be significantly enhanced through changes in these scenarios. Consequently, it is challenging to reduce the energy cost of neutral species (associated with medium-energy (10-20 eV) electrons) simply by increasing the number of harmonics $N$.

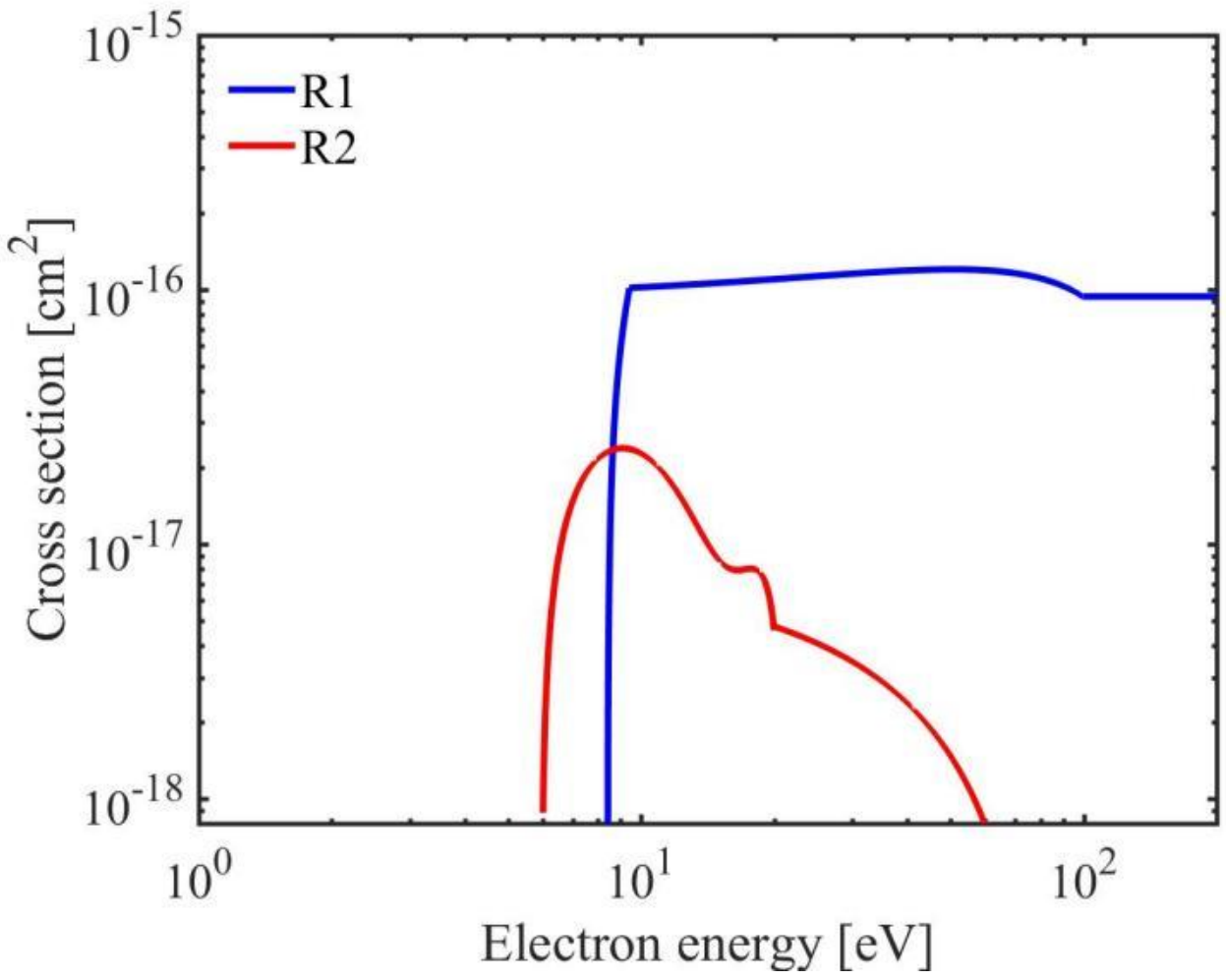

**Figure 6** Collision reaction cross sections between electrons and $O_2$ molecules in the main generation reactions of metastable atom $O(^1D)$ and ground state atom $O(^3P)$ (R1: e + $O_2 \rightarrow O(^3P)$ + $O(^1D)$ + e, R2: e + $O_2 \rightarrow O(^3P) + O(^3P)$ + e).

For ground state atoms $O(^3P)$, their generation mainly depends on electron-induced dissociative collisions reactions (R1 and R2) as well as neutral-neutral collisions reactions (R3 and R4). From figure 6, reaction R2 mainly relies on electrons in the energy range of 8-15 eV. Similar to R1, the occurrence of reaction R2 is predominantly governed by electrons in the medium-energy range (8-20 eV). Thus, as shown in table 1 and table 2, the variation trend of the consecutive relative growth rate of $k$(R2) is consistent with that of $k$(R1). Specifically, as the number of harmonics $N$ increases, under the condition of $r$ = 10%, the consecutive relative growth rate of $k$(R2) changes from −4.73% to −1.4%. And, at $r$ = 50%, the consecutive relative growth rate of $k$(R2) increases from −14.45% to −3.28%. Besides, reactions R3 and R4 exhibit only a weak regulatory effect as a function of $N$, and their rate coefficients are constant [32], with values of $2.1\times10^{-10}$ $cm^3s^{-1}$ and $4.8\times10^{-12}$ $cm^3s^{-1}$, respectively. Therefore, the variation in the total generation rate of $O(^3P)$ ,Gen_$O(^3P)$, will mainly depend on the changing trends of reactions R1 and R2. Besides, the analysis of the electron density has already been discussed in detail. This causes that at $r$ = 10%, the average consecutive relative growth rate of Gen_$O(^3P)$ is 6.25%, with a value of 5.81% at $N$ = 2; and, at $r$ = 50%, the average consecutive relative growth rate Gen_$O(^3P)$ is 1.71%, while it drops to −10.55% at $N$ = 2. It can be seen that the variation trend of the consecutive relative growth rate of the total $O(^3P)$ generation rate is also similar to that of $O(^1D)$.

Further, the variation trend of the energy cost of ground-state $O(^3P)$ atom generation follows the same trend as that of metastable $O(^1D)$ atom (see figure 2). Therefore, the physical mechanism underlying the variation in the generation energy cost of $O(^3P)$ under different $N$ and $r$ can be referred to the previous discussion on $O(^1D)$ and will not be repeated. Moreover, it is can be found that the generation energy cost of ground state atom $O(^3P)$ generation is lower than that of metastable atom $O(^1D)$. This is due to the lower reaction thresholds energy, more generation reactions considered for $O(^3P)$.

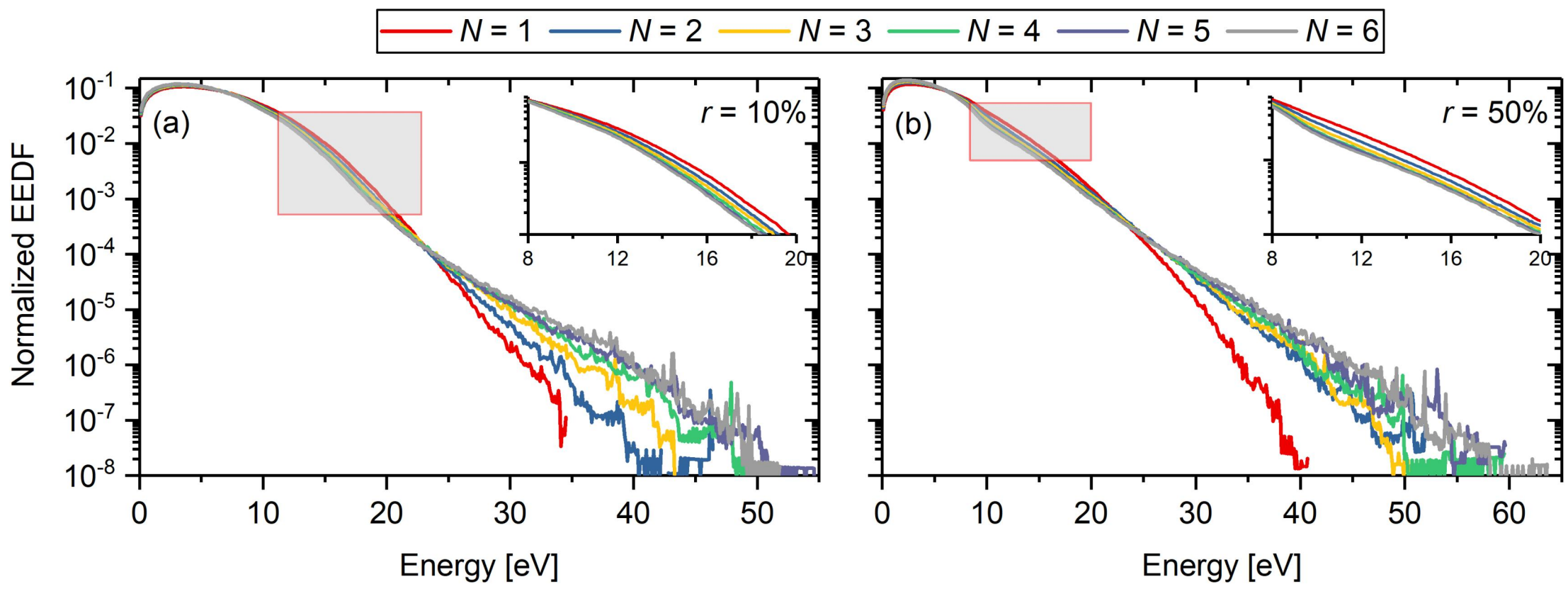

**Figure 7** Spatially and temporally averaged EEDF at different number of harmonics ($N$ = 1-6) for $O_2$ ratio $r$ = 10% (a) and 50% (b). The discharge conditions are consistent with those shown in figure 2.

**Table 1** Consecutive relative growth rates of spatially and temporally averaged electron density $n_e$, rate coefficient $k$(R1) of reaction R1, and rate coefficient $k$(R2) of reaction R2 at $O_2$ ratio ($r$ = 10%) and number of harmonics ($N$ = 1-6).

| $N$ | $\frac{n_{e,N}-n_{e,N-1}}{n_{e,N-1}}$ (%) | $\frac{k(R1)_N-k(R1)_{N-1}}{k(R1)_{N-1}}$ (%) | $\frac{k(R2)_N-k(R2)_{N-1}}{k(R2)_{N-1}}$ (%) |
|---|---|---|---|
| 1 | - | - | - |
| 2 | 18.60 % | -8.74 % | -4.73 % |
| 3 | 17.12 % | -6.32 % | -3.30 % |
| 4 | 14.38 % | -5.45 % | -2.94 % |
| 5 | 11.38 % | -3.37 % | -2.07 % |
| 6 | 8.54 % | -2.56 % | -1.40 % |

**Table 2** Consecutive relative growth rates of spatially and temporally averaged electron density $n_e$, rate coefficient $k$(R1) of reaction R1, and rate coefficient $k$(R2) of reaction R2 at $O_2$ ratio ($r$ = 50%) and number of harmonics ($N$ = 1-6).

| $N$ | $\frac{n_{e,N}-n_{e,N-1}}{n_{e,N-1}}$ (%) | $\frac{k(R1)_N-k(R1)_{N-1}}{k(R1)_{N-1}}$ (%) | $\frac{k(R2)_N-k(R2)_{N-1}}{k(R2)_{N-1}}$ (%) |
|---|---|---|---|
| 1 | - | - | - |
| 2 | 9.25 % | -21.41 % | -14.45 % |
| 3 | 11.99 % | -13.94 % | -10.18 % |
| 4 | 11.22 % | -6.59 % | -4.86 % |
| 5 | 8.97 % | -4.86 % | -3.39 % |
| 6 | 7.43 % | -5.57 % | -3.28 % |

## 4 Conclusions

Based on a one-dimensional (1D) fluid/electron Monte Carlo hybrid model, this work systematically investigates the discharge characteristics of capacitively coupled plasmas operated in Ar/$O_2$ gas mixtures driven by sawtooth up-type voltage waveforms, with particular emphasis on the generation energy cost associated with key neutral species, O($^1$D) and O($^3$P) under different number of harmonics $N$ and $O_2$ ratios.

At a low $O_2$ ratio (10%), the generation energy cost of O($^1$D) and O($^3$P) first decreases and then increases with the increasing number of harmonics $N$, reaching a minimum at $N$ = 2. When the number of harmonics $N$ increases from $N$ = 1 to 2, the total power absorption rate increases slightly. Meanwhile, the generation rates of O($^1$D) and O($^3$P) also increase, and play a dominant role. As a result, the generation energy cost of O($^1$D) and O($^3$P) reaches a minimum at $N$ = 2. The increase in the generation rates of O($^1$D) and O($^3$P) is weakly correlated with the constant background gas density and the decrease in reaction rate coefficients, but dominated by the significant enhancement in electron density. For the reaction rate coefficients of O($^1$D) and O($^3$P), which predominantly depends on medium-energy electrons (approximately 8-20 eV). However, as the number of harmonics $N$ increases, the proportion of high-energy electrons (>20 eV) increases, while the proportion of medium-energy electrons decreases. This leads to a negative growth rate of the reaction rate coefficients. However, for electron density, when the number of harmonics increases from $N$ = 1 to 2, the discharge mode transitions from the DA mode to a hybrid α–DA mode. This process not only enhances the ionization peak but also causes ionization peak to appear near the sheath expansion and collapse phases. In other words, the effective temporal and spatial region of ionization increases, leading to a pronounced increase in electron density.

Besides, as the number of harmonics $N$ further increases, the peak of the ionization rate appears only during the sheath expansion phase, and its enhancement is not significant, resulting in a slower increase in electron density. Although both the total generation rates of O($^1$D) and O($^3$P) and the total power absorption rate show increasing trends during this process, the variation in the total generation rates of O($^1$D) and O($^3$P) is not pronounced. Consequently, the generation energy cost of O($^1$D) and O($^3$P) gradually increases.

As the $O_2$ ratio increases, the discharge operates under highly electronegative conditions. When the number of harmonics $N$ increases from $N$ = 1 to 2 and 3, the effective spatio-temporal region of the generation rates of O($^1$D) and O($^3$P) is reduced due to the modulation of waveform, and meanwhile the enhancement of their peak is limited. Consequently, the generation rates of O($^1$D) and O($^3$P) decrease. As the number of harmonics $N$ is further increased, the effective spatio-temporal region of the generation rates of O($^1$D) and O($^3$P) remains nearly unchanged while their peak values increase, leading to a increase in their total generation rates. Meanwhile, the total power absorption slightly decreases at $N = 2$, which is related to the reduction in electron density caused by the decreased effective spatio-temporal region of the ionization rate. Moreover, during the increase in the number of harmonics $N$, the hybrid α–DA mode does not appear. Therefore, the pronounced enhancement of electron density cannot be observed. These above factors collectively result in a gradual increase in the generation energy cost of $O(^1D)$ and $O(^3P)$ with the increasing number of harmonic $N$, and the inflection point vanishes. However, it can be anticipated that further increasing $N$ accelerates the sheath expansion velocity near the powered electrode, leading to a gradual enhancement of the local ionization rate peak. This may promote the transition to the hybrid α–DA mode, thereby lowering the generation energy cost.

Moreover, the energy cost for the generation of $O(^3P)$ is lower than that for $O(^1D)$, primarily because the total generation rate of $O(^3P)$ is higher than that of $O(^1D)$. Furthermore, the energy cost for the generation of $O(^1D)$ and $O(^3P)$ decreases with increasing $O_2$ ratios. This is because a higher $O_2$ ratio results in a greater $O_2$ density, which in turn enhances the total generation rates of these species and lower energy cost for their generation.

In summary, given the diverse requirements for $O_2$ ratios in practical processes [10, 12, 14], monotonically adjusting the number of harmonic $N$ is insufficient to reduce the energy cost of particles (associated with medium-energy electrons) generation. However, only by selecting the appropriate number of harmonic $N$, the discharge can be effectively sustained in the hybrid α–DA mode, where both the effective spatio-temporal region of the ionization rate and its peak value are enhanced. This, in turn, causes lower energy cost of key neutral species generation. These findings provide meaningful guidance for industrial applications seeking to improve generation energy efficiency. Future research can be further extended to wider pressure ranges, complex process gas systems (e.g., $Ar/O_2/CF_4$, etc.), and different voltage source waveforms.

## Acknowledgments

This work was supported by the National Key R&D Program of China, Grant No. 2025YFF0512000, and the National Natural Science Foundation of China under Grant Nos. 12475202, 12405289.

## Data availability statement

All data that support the findings of this study are included within the article (and any supplementary files).

## ORCID iDs

Jun-Xi Guo https://orcid.org/0009-0004-0182-9328
Wan Dong https://orcid.org/0000-0002-0724-1697
Yuan-Hong Song https://orcid.org/0000-0001-5712-9241